# GHOST Commissioning Science Results III: Characterizing an iron-poor damped Lyman α system

Trystyn A. M. Berg,[1,2,3]⋆ Christian R. Hayes,[1] Stefano Cristiani,[4,5,6] Alan McConnachie,[1] J. Gordon Robertson,[7,8] Federico Sestito,[9] Chris Simpson,[10] Fletcher Waller,[9] Timothy Chin,[7] Adam Densmore,[1] Ruben J. Diaz,[11] Michael L. Edgar,[12] Javier Fuentes Lettura,[2] Manuel Gómez-Jiménez,[11] Venu M. Kalari,[11] Jon Lawrence,[7] Steven Margheim,[13] John Pazder,[1,9] Roque Ruiz-Carmona,[11] Ricardo Salinas,[14] Karleyne M. G. Silva,[11] Katherine Silversides,[1] Kim A. Venn[9]

[1] *NRC Herzberg Astronomy and Astrophysics Research Centre, 5071 West Saanich Road, Victoria, B.C., Canada, V9E 2E7*
[2] *European Southern Observatory, Alonso de Córdova 3107, Vitacura, Santiago, Chile*
[3] *Dipartimento di Fisica G. Occhialini, Università degli Studi di Milano Bicocca, Piazza della Scienza 3, 20126 Milano, Italy*
[4] *INAF−Osservatorio Astronomico di Trieste, Via G.B. Tiepolo, 11, I-34143 Trieste, Italy*
[5] *INFN-National Institute for Nuclear Physics, via Valerio 2, I-34127 Trieste, Italy*
[6] *IFPU−Institute for Fundamental Physics of the Universe, via Beirut 2, I-34151 Trieste, Italy*
[7] *Australian Astronomical Optics - Macquarie University, 105 Delhi Road, North Ryde, NSW 2113, Australia*
[8] *Sydney Institute for Astronomy, School of Physics, University of Sydney, NSW 2006, Australia*
[9] *Department of Physics and Astronomy, University of Victoria, PO Box 3055, STN CSC, Victoria BC V8W 3P6, Canada*
[10] *Gemini Observatory/NSF's NOIRLab, 670 N. A'Ohoku Pl., Hilo, HI 96720, USA*
[11] *Gemini Observatory/NSF's NOIRLab, Casilla 603, La Serena, Chile*
[12] *Anglo Australian Observatory, Australia*
[13] *Vera C. Rubin Observatory/NSF's NOIRLab, Casilla 603, La Serena, Chile*
[14] *Departamento de Física y Astronomía, Universidad de La Serena, Av. Juan Cisternas 1200 Norte, La Serena, Chile*

17 April 2024

**ABSTRACT**
The Gemini High-resolution Optical SpecTrograph (GHOST) is a new echelle spectrograph available on the Gemini-South telescope as of Semester 2024A. We present the first high resolution spectrum of the quasar J1449−1227 (redshift $z_{em} = 3.27$) using data taken during the commissioning of GHOST. The observed quasar hosts an intervening iron-poor ([Fe/H] = −2.5) damped Lyman α (DLA) system at redshift $z = 2.904$. Taking advantage of the high spectral resolving power of GHOST ($R ≈ 55000$), we are able to accurately model the metal absorption lines of the metal-poor DLA and find a supersolar [Si/Fe], suggesting the DLA gas is in an early stage of chemical enrichment. Using simple ionization models, we find that the large range in the C IV/Si IV column density ratio of individual components within the DLA's high ionization absorption profile can be reproduced by several metal-poor Lyman limit systems surrounding the low-ionization gas of the DLA. It is possible that this metal-poor DLA resides within a complex system of metal-poor galaxies or filaments with inflowing gas. The high spectral resolution, wavelength coverage and sensitivity of GHOST makes it an ideal spectrograph for characterizing the chemistry and kinematics of quasar absorption lines.

## 1 INTRODUCTION

The cosmic baryon cycle is a critical process in galaxy evolution. From accreting gas from the intergalactic medium which eventually fuels star formation in the interstellar medium (ISM), to the negative feedback processes that inject energy into the gas to prevent future star formation, the baryon cycle touches many key astrophysical processes that are modestly understood (Tumlinson et al. 2017; Péroux & Howk 2020). Thus understanding how galactic gas reservoirs evolve with cosmic time is an important facet for understanding the rise and fall in cosmic star formation (Madau & Dickinson 2014).

Absorption lines imprinted on top of quasar (QSO) spectra are excellent tracers of galactic gas reservoirs throughout cosmic time. In particular damped Lyα systems (DLAs; Wolfe et al. 2005), with H I column densities N ≥ 2 × 10²⁰ atoms cm⁻², are a key class of QSO absorbers that are expected to trace the main sites of the baryon cycle – the ISM and circumgalactic medium (CGM) of galaxies. In particular, both simulations (Rahmati et al. 2013; Vogelsberger et al. 2014; Rahmati et al. 2015) and observations (Noterdaeme et al. 2009;

⋆ E-mail: Trystyn.Berg@nrc-cnrc.gc.ca

Sánchez-Ramírez et al. 2016) have demonstrated that over $\gtrsim 80\%$ of the neutral gas reservoirs of the Universe reside in DLAs.

Owing to their large column densities, DLAs are also excellent tracers of chemical evolution within galaxies across cosmic time. In terms of bulk chemical enrichment of galactic gas reservoirs, DLAs show a gradual increase in metallicity with time (Pettini et al. 1999; Rafelski et al. 2012; Jorgenson et al. 2013) placing important constraints on cosmological simulations of galaxy evolution (e.g. Hassan et al. 2020; Yates et al. 2021). The advent of high resolution, echelle spectrographs on 8–10m class telescopes (e.g. VLT/UVES, Keck/HIRES, Magellan/MIKE) has enabled detailed chemical abundance studies of individual elements in DLAs, providing a complementary approach to stellar abundance patterns for constraining chemical evolution yields. In particular, both the most metal-poor (with metallicities [M/H]$\lesssim -2.5$) and metal-strong DLAs (column densities logN(Si II)$\geq 15.95$ log(cm$^{-2}$) or logN(Zn II)$\geq 13.15$ log(cm$^{-2}$)) have been used for direct comparison to stellar abundances or constraining stellar yields from the first stars (e.g. Pettini et al. 1997; Prochaska & Wolfe 2002; Cooke et al. 2011; Berg et al. 2015; Welsh et al. 2023).

The Gemini High-resolution Optical SpecTrograph (GHOST; McConnachie et al. 2024, V. Kalari et al. in prep.), is a recently commissioned echelle spectrograph on the Gemini-South telescope. With two integral field unit (IFU) fibre bundles (1.2" in size), GHOST is able to observe two targets simultaneously, or have one IFU for sky measurements. The spectrograph has both a standard and high resolution mode (spectral resolving powers $R \approx 55000$ and $R \approx 76000$; respectively), and provides continuous wavelength coverage from 3475Å to 10610Å. So far GHOST has been used for the detailed chemical abundance analysis of metal-poor stars in the Milky Way (Dovgal et al. 2024; Placco et al. 2023; Sestito et al. 2024) and in Reticulum II ultra-faint dwarf galaxy (Hayes et al. 2023). These works have shown that the instrument has a high efficiency in the bluest regions of the spectrum, hence providing exquisite detection of atomic species' lines that are tracers of the nucleosynthesis in the first generations of stars and their supernovae.

In this paper we present the first observations of a high redshift QSO observed with GHOST in order to test the feasibility of the new Gemini South spectrograph to be used for chemical abundance analyses of QSO absorption lines. Throughout the paper, we assume an Asplund et al. (2009) solar abundance scale.

## 2 TARGET SELECTION, OBSERVATIONS AND DATA REDUCTION

The target QSO J1449−1227 (J2000.0 R.A. 14h:49m:43.17s; Declination −12.0°:27':17.5") was selected from the QUBRICS survey (Calderone et al. 2019; Boutsia et al. 2020; Guarneri et al. 2021) as a potentially interesting candidate for high spectral resolution follow-up. The selected target is a bright (Gaia G = 17.05; Table 1) QSO, previously unobserved at high spectral resolution, with a potential strong H I absorber observed in the QUBRICS confirmation low resolution data (Boutsia et al. 2020). The redshift of J1449−1227 is $z_{\rm em} = 3.27$ (Boutsia et al. 2020).

The QSO J1449−1227 was observed with GHOST during the commissioning of the spectrograph on July 1 2022 from 02:18 UT until 03:48 UT at an airmass of $\approx 1.2$. Conditions were clear, and the average seeing was 1.1". Three consecutive 30-minute exposures were taken, with a total on-target exposure time of 5400s. We used the standard resolution mode, which provides a spectral resolving power of $R \approx 55000$ across the entire wavelength range of both detectors ($\approx 3470$Å to $\approx 10600$Å), and the 2 spectral by 8 spatial pixel binning mode to reduce the effects of read noise.

| Filter | magnitude |
|---|---|
| G[a] | $17.05 \pm 0.004$ |
| g[b] | $17.74 \pm 0.012$ |
| r[b] | $17.09 \pm 0.020$ |
| i[b] | $16.80 \pm 0.009$ |
| z[b] | $16.69 \pm 0.005$ |
| y[b] | $16.64 \pm 0.013$ |

[a] GAIA DR3 (Gaia Collaboration et al. 2023)
[b] PAN-STARRS DR1 (Chambers et al. 2016)

**Table 1.** QSO J1449−1227 apparent magnitudes

The observed data were reduced using the GHOSTDR data reduction pipeline (Ireland et al. 2018; Hayes et al. 2022)[1], which performs the optimal extraction, wavelength calibration, blaze correction, and order stitches the spectra to produce 1D extracted spectra for GHOST's red and blue science cameras. Given the stability of the instrument, we note that bias and flats were taken during the week of the commissioning run, and the arc calibrations were taken the evening preceding the observation. No dark subtraction was performed given the very low dark current of the GHOST spectrograph (McConnachie et al. 2024). After correcting for barycentric velocity shifts, the three individual-exposure spectra, and the corresponding error spectra, were combined by taking the mean spectrum. A first-order QSO continuum was fitted to each chip's spectrum using a cubic spline, where the spline knots were visually selected as pixels that are expected to contain pure continuum.

Figure 1 shows the relative flux (not flux-calibrated and in arbitrary units; top panel) and resulting signal-noise ratio (S/N; bottom panel) of the combined spectrum prior to continuum normalization. While we do not have a comparison spectrum taken with other instruments, we note that the S/N obtained by GHOST is comparable to the expected S/N[2] from VLT/UVES (Dekker et al. 2000) in similar observing conditions using a 0.8" slit ($R \approx 55000$). While the two instruments appear comparable, GHOST is able to obtain the full wavelength coverage ($\approx 3470$Å to 10600Å) in one shot, whereas UVES would require at least two different set-ups to get a similar wavelength coverage.

## 3 DATA ANALYSIS AND DISCUSSION

### 3.1 Characterizing the intervening DLA

#### 3.1.1 The H I column density of the strong absorber

Figure 2 shows the strong Ly$\alpha$ absorption feature observed in the GHOST spectrum. The H I column density of the absorber was determined by simultaneously fitting the continuum along with a single Voigt profile to the Ly$\alpha$ absorption in order to reproduce the core and wings of the absorption profile. The resulting H I column density of the absorber is logN(H I) = $20.55 \pm 0.15$ log(cm$^{-2}$) at the redshift $z = 2.9049$. The errors on logN(H I) were determined visually by encompassing the variation in the flux within the Voigt profile's wings.

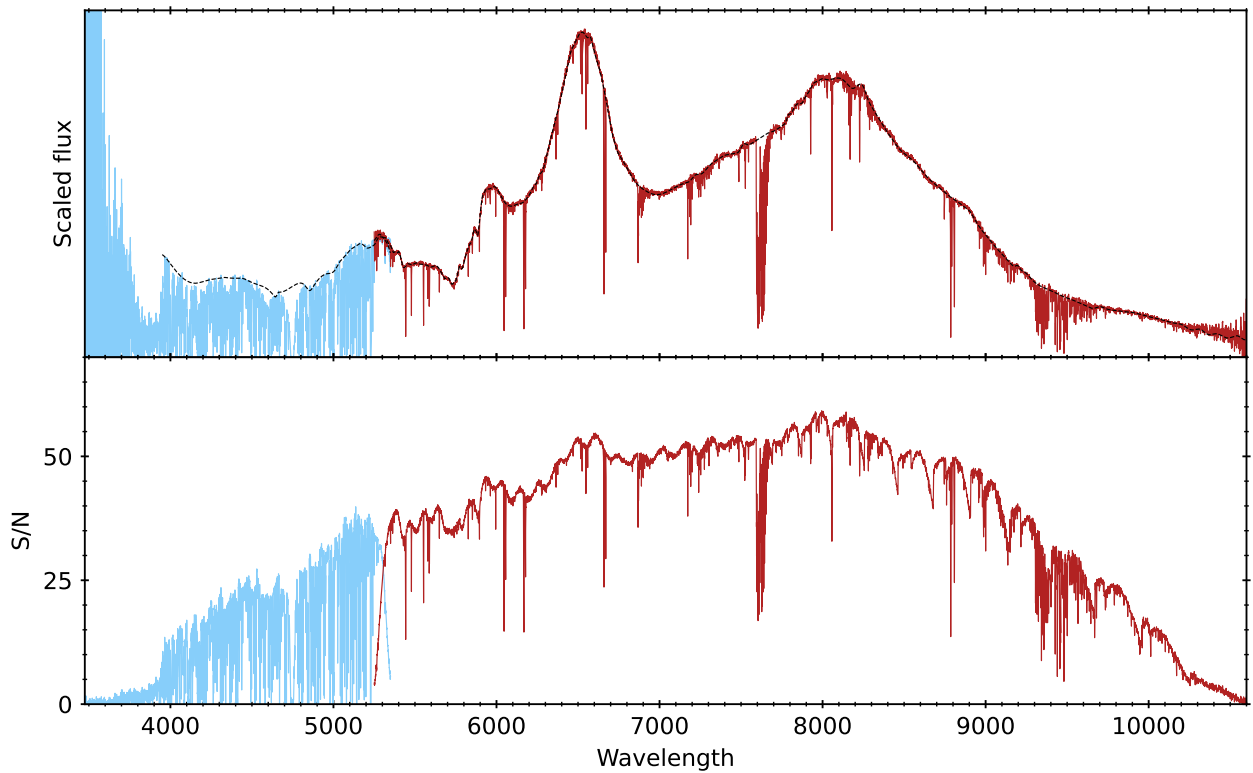

**Figure 1.** The full spectrum (top panel) and estimated S/N (bottom panel) of the obtained GHOST spectrum for QSO J1449−1227. The curves in both samples have been rebinned by 10 pixels for display purposes. The light blue and dark red lines denote the spectra obtained from the respective blue and red chip. The dashed black line in the top panel shows the first-order continuum fit used in the analysis. We note that the Lyα emission of the QSO is located within the overlapping wavelength region of the two chips ($\approx 5300\text{Å}$). The spectra in the top panel have not been flux-corrected and are in arbitrary flux units. Both the red and blue chip spectra in the top panel have been normalized to match the flux in the overlapping wavelength region at $\approx 5300\text{Å}$ for display purposes.

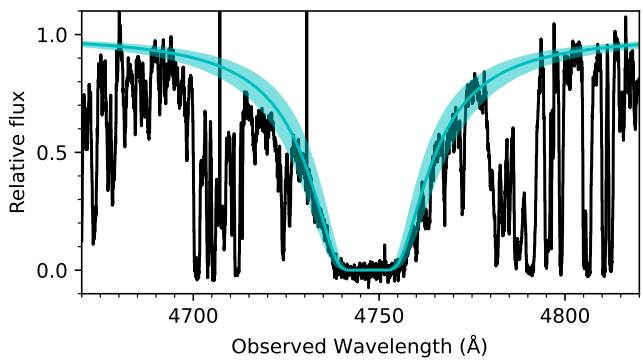

**Figure 2.** The strong Lyα absorption observed in the spectrum of J1449−1227 (black line). The cyan curve represents the best-fit Voigt profile to the DLA absorber with an H I column density of logN(H I) = $20.55 \pm 0.15$ log(cm$^{-2}$). The shaded cyan region demonstrates the resulting error in the Voigt profile from the adopted error on the H I column density.

### 3.1.2 Metal line fitting

Metal absorption lines associated with the DLA were visually identified then fitted using ALIS[3]. ALIS uses a least-squares fitting approach to simultaneously model the QSO continuum, absorption line components (characterized by a redshift $z_{\rm comp}$, turbulent Doppler broadening parameter $b_{\rm comp}$, column density logN$_{\rm comp}$(X), and gas temperature $T_{gas}$), and blends. We assume a fixed gas temperature typical for metal-poor DLAs of $T_{gas} = 10^4 K$ (Cooke et al. 2015) as we do not have the range in atomic mass and S/N to accurately decouple the gas temperature from turbulent broadening. Therefore, the measured $b_{\rm comp}$ does not fully represent the turbulent component of the gas. We assume Gaussian instrumental broadening with a full-width half maximum (FWHM) of 5.3 km s$^{-1}$ for the blue arm and 5.2 km s$^{-1}$ for the red arm. A 3$^{\rm rd}$ order Legendre polynomial was fitted to the continuum in order to pick up any minimal curvature not fully accounted for in the first-order continuum fits.

[3] https://github.com/rcooke-ast/ALIS

For components that are common between different velocity profiles of the same ionization state, we assume the same $z_{\rm comp}$ and $b_{\rm comp}$ and only vary logN between different ions. Figures 3 and 4 show the DLA's detected metal lines (black lines) for the low and high ionization species, respectively. For each specific velocity profile, we only include components in the final model (magenta lines in Figures 3 and 4) that are detected at $\geq 1\sigma$ confidence. Where there is nearby blending, additional independent components (cyan lines) were added in order to ensure that the continuum is well modelled. The residual in the relative flux of the data relative to the profile fits is denoted by the dark green lines in Figures 3 and 4. The residual is scaled on a pixel-by-pixel basis using the error spectrum such that the top and bottom of the horizontal grey band represents the data deviating from the model profile by $\pm 1\sigma$ respectively. For metal lines that contain significant blending that makes the fitting difficult (i.e. C ɪɪ 1036Å, Si ɪɪ 1260Å, Si ɪɪ 1304Å, Fe ɪɪ 1608Å[4], Fe ɪɪ 2249Å, and Fe ɪɪ 2344Å), we show the expected modelled line profiles (red line) based on the fits of the other species' line(s). As we are limited to only having O ɪ 1302Å for this absorber, we continue to use this O ɪ line despite the significant amount of blending. However, we conservatively assume the obtained column density is a lower limit due to both the saturation and spectral blending that prevents accurate fitting of the component at $z_{comp}$ = 2.90490.

### 3.1.3 Chemical abundance profile

As the bulk of the gas is assumed to be in the neutral phase in DLAs due to self-shielding from H ɪ, we compute the metallicity and chemical abundance profile using the column densities of the low-ionization species relative to logN(H ɪ) (Wolfe et al. 2005). Table 4 presents the total column densities and abundance profiles of this DLA. The total column densities of C ɪɪ and O ɪ are conservatively set as lower limits due to the line profiles being slightly saturated, in addition to O ɪ being significantly blended. The chemical abundance profile of this DLA is a bit peculiar. While the observed [Si/H] ($-2.14 \pm 0.15$) is typical for DLAs at this redshift (Rafelski et al. 2012; Berg et al. 2015), the measured [Fe/H] = $-2.55 \pm 0.15$ makes this one of the most Fe-poor DLAs identified to date (Welsh et al. 2022). While we do not have sufficient abundance measurements for different species to accurately assess dust depletion using common nucleosynthetic tracers (Berg et al. 2016), we expect dust depletion for this [Fe/H] to be minimal based on the dust depletion sequences of other Fe-poor DLAs (De Cia et al. 2018). If there is indeed no dust depletion in the DLA towards QSO J1449−1227, the supersolar [Si/Fe] = $0.41 \pm 0.03$ must be primarily driven by nucleosynthesis suggesting that the gas in this absorber has not had significant enrichment from Type ɪa supernovae. Furthermore, the enhanced [C/Fe] ($> 0.17$) suggests there is an enhancement in the carbon abundance of this absorber that is consistent with other metal-poor DLAs (Welsh et al. 2022). The subsolar [Al/Si] = $-0.29 \pm 0.02$ is consistent with other metal-poor DLAs showing a strong odd-even effect (Cooke et al. 2017).

### 3.1.4 Cloudy modelling of high ionization gas

The range of N(C ɪᴠ)/N(Si ɪᴠ) seen in the components of the high ionization lines associated with the DLA ($\approx$ 1.8 − 16; Table 3) suggest the neutral gas is embedded in a complex environment of different density clouds. We ran a simple grid of Cʟᴏᴜᴅʏ (Ferland et al. 2017, version 17.03) photoionization models in order to estimate the physical properties (i.e. metallicity [M/H], volume density $n_H$, and logN(H ɪ)) of each individual component of the high ion absorption profile based on the measured logN(C ɪᴠ) and logN(Si ɪᴠ). In the Cʟᴏᴜᴅʏ grid, we assume a slab of plane-parallel gas illuminated by the Haardt & Madau (2012) UV background at the redshift of the absorber ($z = 2.90522$) on the solar abundance scale. The grid was run by varying [M/H], $n_H$, and total logN(H ɪ) for the range of values listed in Table 5. We note that the maximum logN(H ɪ) of the Cʟᴏᴜᴅʏ grid is set by the observed H ɪ column density of the DLA. The grid was then interpolated to a finer resolution (Table 5) using a linear grid interpolation. We find the best-fitting set of interpolated Cʟᴏᴜᴅʏ model parameters for each absorption component that simultaneously matches the observed N(C ɪᴠ) and N(Si ɪᴠ) within the measured $1\sigma$ errors. Table 6 shows the best matching Cʟᴏᴜᴅʏ model parameters (i.e. minimal offset), along with the range of parameter values that are consistent with the column densities within the measured $1\sigma$ errors in N(C ɪᴠ) and N(Si ɪᴠ). Following Saccardi et al. (2023), we repeated the same experiment using the abundance pattern of metal-poor stars in Cayrel et al. (2004) in place of a solar pattern within Cʟᴏᴜᴅʏ. The Cayrel et al. (2004) scale reproduces the same bulk trends in the best-fitting model parameters shown Table 6 as the solar abundance scale. While the model parameters $n_H$ and logN(H ɪ) resulting from the two different abundance patterns are consistent within the $1\sigma$ errors, the best-fitting [M/H] are systematically $\approx$ 0.8 dex lower for each component using the Cayrel et al. (2004) scale.. Despite the fitting procedure being unconstrained, the range of high N(H ɪ) and typically low [M/H] of these components suggests the C ɪᴠ and Si ɪᴠ are associated with several metal-poor Lyman limit systems (LLS; logN(H ɪ) $\geq$ 17) surrounding the neutral DLA gas. Based on observations of direct detection of galaxies associated with LLS gas, it is possible that this high-ionization gas is associated with several nearby galaxies (Lofthouse et al. 2023), or inflowing metal-poor gas (Fumagalli et al. 2016).

As C ɪᴠ and Si ɪᴠ can trace both photo- and collisionally ionized gas (e.g. Fox et al. 2007b; Werk et al. 2019; Hasan et al. 2022), more complex ionization modelling is likely required in order to fully constrain the physical properties of the gas clouds associated with each component of the absorption profile. Unfortunately with only C ɪᴠ and Si ɪᴠ, we are limited in our ability to simultaneously constrain the many physical parameters required in full ionization modelling. Collisional ionization is expected to be present in DLA gas associated with C ɪᴠ and Si ɪᴠ when the gas temperature is above $T_{gas} \gtrsim 70\,000K$ (Fox et al. 2007a). While it is possible to model $T_{gas}$ in the line fitting process, we are unable to obtain accurate $T_{gas}$ measurements for each component of the C ɪᴠ and Si ɪᴠ profiles with this dataset. Including $T_{gas}$ as a free parameter in the line fitting process results in unconstrained bounds on $T_{gas}$ for each component (e.g. $T_{gas}$ is consistent with = 0 K). Higher spectral resolution ($R \approx 150\,000$) with improved signal-noise may enable a more accurate decoupling of the turbulent and thermal broadening components in the line profile.

In an attempt to obtain an estimated range in $T_{gas}$, we repeated the component fitting models for the high ions using $T_{gas}$ as a free parameter whilst fixing $b_{comp}$ to 1.0 km s$^{-1}$ (i.e. minimal turbulence) and 8.0 km s$^{-1}$ (maximal turbulence; consistent with the low ionization gas) as upper and lower limits to the amount of thermal broadening[5]. Figure 5 shows the resulting logN(C ɪᴠ)/logN(Si ɪᴠ) for each component of the C ɪᴠ and Si ɪᴠ profiles as a function of the best-fit $T_{gas}$. For comparison, the predicted logN(C ɪᴠ)/logN(Si ɪᴠ) as a function of $T_{gas}$ resulting from collisional ionization equilibrium (Gnat & Sternberg 2007) is shown as the black line. Based on modelling of the Milky Way's CGM (Werk et al. 2019), it is expected that larger values of logN(C ɪᴠ)/logN(Si ɪᴠ) inconsistent with ratios predicted from pure collisional ionization tend to require a mix of both photo- and collisional ionization. For at least seven of the components in the DLA's high ionization absorption profile, there is likely a mix of both collisional and photo-ionization processes using either the minimum or maximum turbulent Doppler parameter. We thus expect that the best-fit H ɪ column densities obtained from the

---

[4] The strongest component of the Fe ɪɪ 1608Å line falls directly in telluric absorption. As the Fe ɪɪ profile is well constrained from the other Fe ɪɪ lines, Fe ɪɪ 1608 Å is not included in the velocity profile fitting.

[5] We note that, if the true $b_{comp}$ is higher than 8 km s$^{-1}$, the resulting component's $T_{gas}$ will much lower and consistent with a predominately turbulent medium.

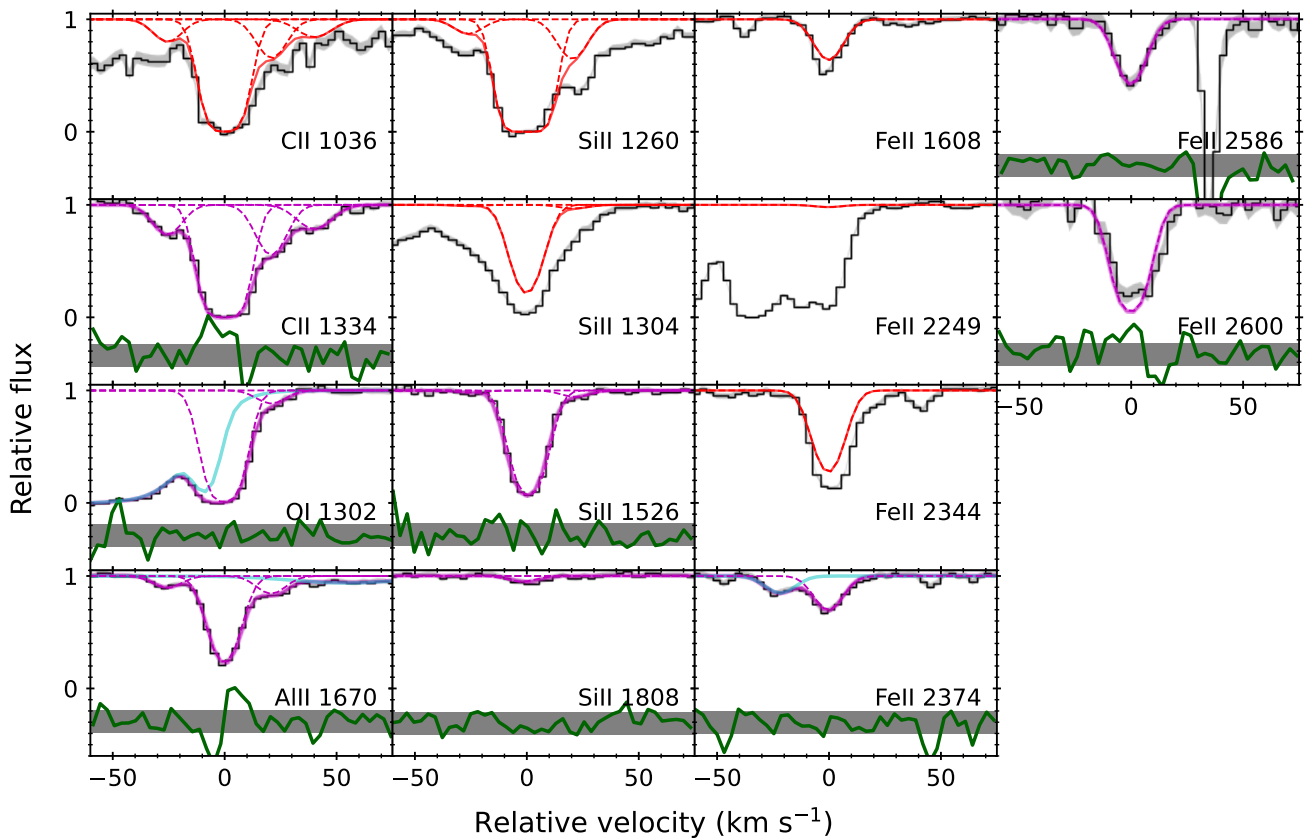

**Figure 3.** The continuum-normalized absorption profiles of the low ionization lines (black lines) observed in the spectrum of J1449−1227. The relative velocity scale is normalized to the redshift of the strongest component of the absorption profile ($z_{\rm comp} = 2.90523$). The light grey shaded regions surrounding the black curve denote the $\pm 1\sigma$ uncertainty range on the flux. The velocity profile fits are shown in magenta (with dashed magenta lines denoting individual components), with added components to fit blends denoted in cyan. The scaled residual from the data with respect to the fits are denoted by the green curve. The scaling of the residual is done for each pixel such that the top and bottom of the horizontal grey bar represents a $\pm 1\sigma$ deviation from the model's relative flux based on the error spectrum. For lines that were not included in the fitting process (C ɪɪ 1036, Si ɪɪ 1304, Fe ɪɪ 1608, Fe ɪɪ 2249, and Fe ɪɪ 2344), the predicted profiles and components are shown in red.

| $z_{\rm comp}$ | $b_{\rm comp}$ km s$^{-1}$ | $\log{\rm N_{comp}}$(C ɪɪ) log(cm$^{-2}$) | $\log{\rm N_{comp}}$(O ɪ) log(cm$^{-2}$) | $\log{\rm N_{comp}}$(Al ɪɪ) log(cm$^{-2}$) | $\log{\rm N_{comp}}$(Si ɪɪ) log(cm$^{-2}$) | $\log{\rm N_{comp}}$(Fe ɪɪ) log(cm$^{-2}$) |
|---|---|---|---|---|---|---|
| 2.90490 | $8.09 \pm 0.74$ | $13.03 \pm 0.03$ | . . . | $11.36 \pm 0.05$ | $11.74 \pm 0.30$ | . . . |
| 2.90523 | $8.00 \pm 0.10$ | $14.55 \pm 0.02$ | $14.67 \pm 0.03$ | $12.47 \pm 0.01$ | $13.90 \pm 0.01$ | $13.45 \pm 0.02$ |
| 2.90550 | $8.22 \pm 0.89$ | $13.32 \pm 0.04$ | $13.07 \pm 0.13$ | $11.55 \pm 0.07$ | $12.24 \pm 0.10$ | . . . |
| 2.90575 | $9.93 \pm 1.55$ | $13.00 \pm 0.06$ | . . . | . . . | . . . | . . . |

**Table 2.** Voigt profile fitting parameters for low ionization species.

Cʟᴏᴜᴅʏ photoionization model of each component are likely close to the true H ɪ column densities of each component, suggesting there are metal-poor LLS clouds surrounding the DLA.

### 3.2 Other absorbing systems

In addition to the DLA clearly identified by the strong Lyα absorption, we also inspected the spectrum for the presence of strong systems characterized by strong C ɪᴠ 1548Å, 1550Å or Mg ɪɪ 2796Å, 2803Å doublets (i.e. where the strongest line has a rest-frame equivalent width EW ≥ 0.1Å). There are three Mg ɪɪ systems (Table 7) and three C ɪᴠ systems (Table 8; excluding the DLA). We note that all three Mg ɪɪ systems are likely strong H ɪ absorbers (i.e. $\log{\rm N(H\,ɪ)} \gtrsim 19.0$ log(cm$^{-2}$)) based on the Mg ɪɪ 2796Å EW cuts proposed by low redshift DLA observations (Rao et al. 2017). The three C ɪᴠ systems appear to be associated with Lyman limit systems ($\log{\rm N(H\,ɪ)} \geq 17.0$ log(cm$^{-2}$)) based on the strength of the Lyα profiles. Unfortunately the absorption blueward of the Lyman limit of the QSO ($\lesssim 3890$ Å) prevents being able to identify the H ɪ column density of the highest redshift Mg ɪɪ system.

## 4 CONCLUSIONS

This paper presents the first high-redshift QSO spectrum obtained by GHOST, the new high-resolution optical spectrograph on Gemini-South. The target, J1449−1227, is a relatively bright QSO ($G = 17.0$) at redshift $z_{\rm em} = 3.27$. During GHOST commissioning, the target QSO was observed with the standard resolution mode using

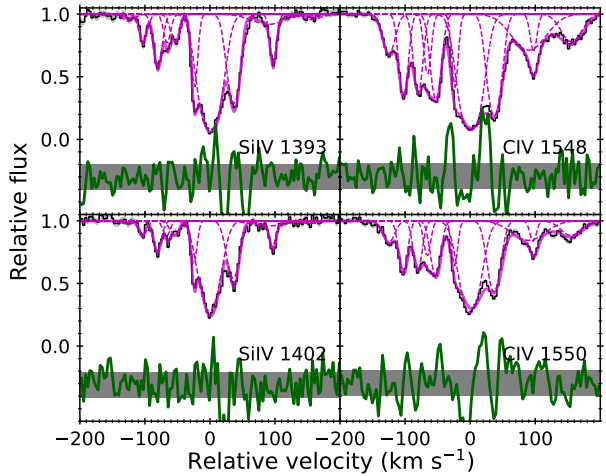

**Figure 4.** The absorption profiles of the high ionization lines for the DLA towards J1449−1227. reference velocity and notation are the same as Figure 3.

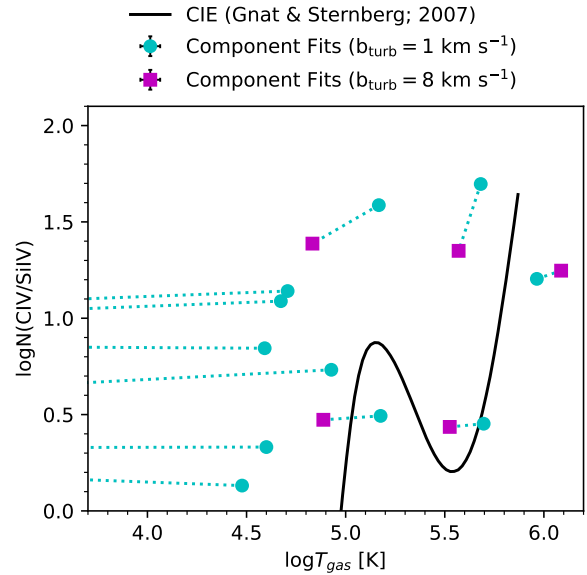

**Figure 5.** The column density ratio of logN(C IV)/logN(Si IV) for each component of the high ionization lines' absorption profiles as a function of the gas temperature ($T_{gas}$). The cyan circles and magenta squares represent the components' profile parameters resulting from assuming a turbulent Doppler broadening parameter of 1 and 8 km s$^{-1}$, respectively. We note the errors on the measurements are smaller than the points. The dotted cyan lines are used to connect the model of the two different turbulence assumptions for the same component. Magenta points not shown in the plot are consistent with a predominantly turbulent medium. The black line shows the expected column density ratio in purely collisionally ionized gas in equilibrium (CIE; Gnat & Sternberg 2007). The degree to which points are offset from the black line is likely a result of the increasing amount of photoionization present in the gas cloud (Werk et al. 2019). At least seven of the velocity profile components appear to be consistent with a having a significant amount of photoionized material.

| $z_{\rm comp}$ | $b_{\rm comp}$ km s$^{-1}$ | logN$_{\rm comp}$(C IV) log(cm$^{-2}$) | logN$_{\rm comp}$(Si IV) log(cm$^{-2}$) |
|---|---|---|---|
| 2.90359 | 11.91 ± 0.83 | 13.04 ± 0.03 | 11.82 ± 0.11 |
| 2.90388 | 8.16 ± 0.34 | 13.34 ± 0.01 | 12.32 ± 0.03 |
| 2.90417 | 8.17 ± 0.58 | 13.31 ± 0.03 | 12.67 ± 0.03 |
| 2.90436 | 6.33 ± 1.06 | 13.04 ± 0.11 | 12.26 ± 0.10 |
| 2.90453 | 9.66 ± 0.64 | 13.43 ± 0.04 | 12.38 ± 0.06 |
| 2.90491 | 5.25 ± 0.36 | 13.11 ± 0.04 | 12.72 ± 0.04 |
| 2.90522 | 19.48 ± 0.41 | 14.08 ± 0.01 | 13.68 ± 0.01 |
| 2.90569 | 12.48 ± 0.25 | 13.69 ± 0.01 | 13.18 ± 0.01 |
| 2.90639 | 35.02 ± 2.22 | 13.43 ± 0.02 | 12.45 ± 0.06 |
| 2.90648 | 5.89 ± 0.39 | 12.75 ± 0.04 | 12.49 ± 0.03 |
| 2.90724 | 20.90 ± 1.28 | 13.08 ± 0.03 | 11.93 ± 0.11 |

**Table 3.** Voigt profile fitting parameters for high ionization species.

| Ion | logN(X) [log(cm$^{-2}$)] | [X/H] | [X/Fe] |
|---|---|---|---|
| C II | > 14.60 | > −2.38 | > 0.17 |
| C IV | 14.50 ± 0.01 | ... | ... |
| O I | > 14.68 | > −2.56 | > −0.01 |
| Al II | 12.55 ± 0.01 | −2.43 ± 0.15 | 0.12 ± 0.03 |
| Si II | 13.92 ± 0.01 | −2.14 ± 0.15 | 0.41 ± 0.03 |
| Si IV | 13.94 ± 0.01 | ... | ... |
| Fe II | 13.45 ± 0.02 | −2.55 ± 0.15 | − |

**Table 4.** Total metal column densities and abundances for the DLA towards QSO J1449−1227.

| Parameter | Range | CLOUDY increment | Interpolation resolution |
|---|---|---|---|
| $n_H$ [cm$^{-3}$] | [−4.0,4.0] | 0.5 | 0.05 |
| logN(H I) [cm$^{-2}$] | [14.0,20.6] | 0.3 | 0.1 |
| [M/H] | [−4.7,−0.2] | 0.5 | 0.05 |

**Table 5.** Range of parameters of the CLOUDY model grid.

2 × 8 spectral by spatial binning. The spectrum reached a typical S/N of 35 − 50 pixel$^{-1}$ redward of the Ly$\alpha$ emission of the quasar for an on-sky exposure time of 5400s (Figure 1). The obtained S/N is comparable to that expected observing the same QSO with UVES (0.8" slit) in similar observing conditions. However, GHOST is able to provide simultaneous wavelength coverage of 3500Å to 10600Å compared to the multiple dichroic set-ups required by UVES to obtain the same wavelength coverage.

An iron-poor DLA ($z_{abs}$ = 2.905, logN(H I) = 20.55 ± 0.15 log(cm$^{-2}$); [Fe/H] = −2.55±0.15) was characterized in the GHOST spectrum. This iron-poor DLA has a similar abundance pattern compared to other metal-poor DLAs in the literature (Cooke et al. 2017; Welsh et al. 2022). Using a simple photoionization model, we find that the high ionization gas traced by C IV and Si IV can be reproduced by several metal-poor LLS surrounding a single DLA cloud, suggesting the DLA resides in a complex environment of other galaxies or inflowing gas (Fumagalli et al. 2016; Lofthouse et al. 2023).

Based on similar literature analyses using other optical high-resolution spectrographs, GHOST is able to provide similar high quality spectra that enable accurate characterization of the kinematics and chemical abundances of quasar absorption lines. The combination of the high spectral resolution ($R \approx 55000$), efficiency and throughput of GHOST makes the new spectrograph an ideal instrument for characterizing the kinematics and chemical abundance profiles of quasar absorption line systems.

| $z_{comp}$ | logN(C IV)/logN(Si IV) | $n_H$ [cm$^{-3}$] | | logN(H I) [cm$^{-2}$] | | [M/H] | |
|---|---|---|---|---|---|---|---|
| | | Best-fit | 1σ-Range | Best-fit | 1σ-Range | Best-fit | 1σ-Range |
| 2.90359 | 1.21 ± 0.013 | −3.7 | [−3.8,−2.9] | 17.5 | [15.1,17.6] | −4.6 | [−4.7,−1.3] |
| 2.90388 | 1.02 ± 0.007 | −3.6 | [−3.7,−2.8] | 18.2 | [15.7,20.6] | −4.5 | [−4.6,−1.3] |
| 2.90417 | 0.64 ± 0.008 | −2.7 | [−3.0,−2.5] | 17.4 | [16.1,20.6] | −2.7 | [−3.5,−1.3] |
| 2.90436 | 0.78 ± 0.016 | −3.2 | [−3.6,−2.5] | 18.2 | [15.6,20.6] | −4.1 | [−4.7,−1.3] |
| 2.90453 | 1.05 ± 0.010 | −3.7 | [−3.8,−2.8] | 18.3 | [15.7,20.6] | −4.5 | [−4.7,−1.3] |
| 2.90491 | 0.39 ± 0.009 | −2.5 | [−2.6,−2.3] | 18.1 | [16.5,20.6] | −2.7 | [−2.9,−1.3] |
| 2.90522 | 0.40 ± 0.004 | −2.5 | [−2.5,−2.5] | 17.5 | [17.5,17.9] | −1.5 | [−1.8,−1.5] |
| 2.90569 | 0.52 ± 0.005 | −2.5 | [−2.7,−2.5] | 16.7 | [16.7,18.3] | −1.3 | [−2.6,−1.3] |
| 2.90648 | 0.26 ± 0.009 | −2.2 | [−2.4,−2.1] | 17.5 | [16.6,20.6] | −2.3 | [−2.8,−1.3] |
| 2.90639 | 0.99 ± 0.010 | −3.0 | [−3.7,−2.8] | 17.1 | [15.8,20.6] | −2.8 | [−4.5,−1.3] |
| 2.90724 | 1.15 ± 0.013 | −3.5 | [−3.7,−2.8] | 17.3 | [15.2,17.9] | −4.1 | [−4.7,−1.3] |

**Table 6.** Best fit CLOUDY model parameters for high ionization line components.

| $z_{abs}$ | EW$_{2796}$ Å | EW$_{2803}$ Å |
|---|---|---|
| 1.69153 | 0.286 ± 0.006 | 0.099 ± 0.005 |
| 2.14169 | 0.812 ± 0.005 | 0.623 ± 0.005 |
| 2.38131 | 1.503 ± 0.008 | 1.216 ± 0.012 |

**Table 7.** Strong Mg II systems and rest-frame equivalent widths of the doublet lines.

| $z_{abs}$ | EW$_{1548}$ Å | EW$_{1550}$ Å |
|---|---|---|
| 2.40057 | 0.163 ± 0.018 | 0.077 ± 0.013 |
| 2.98390 | 0.632 ± 0.003 | 0.418 ± 0.003 |
| 3.11124 | 0.100 ± 0.003 | 0.043 ± 0.003 |

**Table 8.** Strong C IV systems and rest-frame equivalent widths of the doublet lines.

## ACKNOWLEDGMENTS

We would like to thank the Gemini staff for their help and support in making the GHOST commissioning such a success, and the referee for their useful comments to improve the clarity of the paper. TAMB is grateful for the useful discussions with Ryan Cooke, Valentina D'Odorico, Louise Welsh, and Jessica Werk that helped improve this manuscript. KAV and FW are grateful for funding from the NSERC CREATE and NSERC Discovery grant programs. This paper is based on observations obtained at the international Gemini Observatory, a program of NSF's NOIRLab, which is managed by the Association of Universities for Research in Astronomy (AURA) under a cooperative agreement with the National Science Foundation on behalf of the Gemini Observatory partnership: the National Science Foundation (United States), National Research Council (Canada), Agencia Nacional de Investigación y Desarrollo (Chile), Ministerio de Ciencia, Tecnología e Innovación (Argentina), Ministério da Ciência, Tecnologia, Inovações e Comunicações (Brazil), and Korea Astronomy and Space Science Institute (Republic of Korea).

## DATA AVAILABILITY

The data underlying this article are available in the article and in its online supplementary material. The supplementary material contains the spectrum, error spectrum and fitted continuum.

## REFERENCES

Asplund M., Grevesse N., Sauval A. J., Scott P., 2009, ARA&A, 47, 481
Berg T. A. M., Ellison S. L., Prochaska J. X., Venn K. A., Dessauges-Zavadsky M., 2015, MNRAS, 452, 4326
Berg T. A. M., et al., 2016, MNRAS, 463, 3021
Boutsia K., et al., 2020, ApJS, 250, 26
Calderone G., et al., 2019, ApJ, 887, 268
Cayrel R., et al., 2004, A&A, 416, 1117
Chambers K. C., et al., 2016, arXiv e-prints, p. arXiv:1612.05560
Cooke R., Pettini M., Steidel C. C., Rudie G. C., Nissen P. E., 2011, MNRAS, 417, 1534
Cooke R. J., Pettini M., Jorgenson R. A., 2015, ApJ, 800, 12
Cooke R. J., Pettini M., Steidel C. C., 2017, MNRAS, 467, 802
De Cia A., Ledoux C., Petitjean P., Savaglio S., 2018, A&A, 611, A76
Dekker H., D'Odorico S., Kaufer A., Delabre B., Kotzlowski H., 2000, in Iye M., Moorwood A. F., eds, Society of Photo-Optical Instrumentation Engineers (SPIE) Conference Series Vol. 4008, Optical and IR Telescope Instrumentation and Detectors. pp 534–545, doi:10.1117/12.395512
Dovgal A., et al., 2024, MNRAS, 527, 7810
Ferland G. J., et al., 2017, Rev. Mex. Astron. Astrofis., 53, 385
Fox A. J., Petitjean P., Ledoux C., Srianand R., 2007a, A&A, 465, 171
Fox A. J., Ledoux C., Petitjean P., Srianand R., 2007b, A&A, 473, 791
Fumagalli M., Cantalupo S., Dekel A., Morris S. L., O'Meara J. M., Prochaska J. X., Theuns T., 2016, MNRAS, 462, 1978
Gaia Collaboration et al., 2023, A&A, 674, A1
Gnat O., Sternberg A., 2007, ApJS, 168, 213
Guarneri F., Calderone G., Cristiani S., Fontanot F., Boutsia K., Cupani G., Grazian A., D'Odorico V., 2021, MNRAS, 506, 2471
Haardt F., Madau P., 2012, ApJ, 746, 125
Hasan F., Churchill C. W., Stemock B., Nielsen N. M., Kacprzak G. G., Croom M., Murphy M. T., 2022, ApJ, 924, 12
Hassan S., Finlator K., Davé R., Churchill C. W., Prochaska J. X., 2020, MNRAS, 492, 2835
Hayes C. R., et al., 2022, in Evans C. J., Bryant J. J., Motohara K., eds, Society of Photo-Optical Instrumentation Engineers (SPIE) Conference Series Vol. 12184, Ground-based and Airborne Instrumentation for Astronomy IX. p. 121846H, doi:10.1117/12.2642905
Hayes C. R., et al., 2023, ApJ, 955, 17
Ireland M. J., White M., Bento J. P., Farrell T., Labrie K., Luvaul L., Nielsen J. G., Simpson C., 2018, in Guzman J. C., Ibsen J., eds, Society of Photo-Optical Instrumentation Engineers (SPIE) Conference Series Vol. 10707, Software and Cyberinfrastructure for Astronomy V. p. 1070735, doi:10.1117/12.2314418
Jorgenson R. A., Murphy M. T., Thompson R., 2013, MNRAS, 435, 482
Lofthouse E. K., et al., 2023, MNRAS, 518, 305
Madau P., Dickinson M., 2014, ARA&A, 52, 415
McConnachie A. W., et al., 2024, arXiv e-prints, p. arXiv:2401.07452
Noterdaeme P., Petitjean P., Ledoux C., Srianand R., 2009, A&A, 505, 1087

Péroux C., Howk J. C., 2020, ARA&A, 58, 363

Pettini M., Smith L. J., King D. L., Hunstead R. W., 1997, ApJ, 486, 665

Pettini M., Ellison S. L., Steidel C. C., Bowen D. V., 1999, ApJ, 510, 576

Placco V. M., et al., 2023, ApJ, 959, 60

Prochaska J. X., Wolfe A. M., 2002, ApJ, 566, 68

Rafelski M., Wolfe A. M., Prochaska J. X., Neeleman M., Mendez A. J., 2012, ApJ, 755, 89

Rahmati A., Pawlik A. H., Raičević M., Schaye J., 2013, MNRAS, 430, 2427

Rahmati A., Schaye J., Bower R. G., Crain R. A., Furlong M., Schaller M., Theuns T., 2015, MNRAS, 452, 2034

Rao S. M., Turnshek D. A., Sardane G. M., Monier E. M., 2017, MNRAS, 471, 3428

Saccardi A., et al., 2023, ApJ, 948, 35

Sánchez-Ramírez R., et al., 2016, MNRAS, 456, 4488

Sestito F., et al., 2024, MNRAS,

Tumlinson J., Peeples M. S., Werk J. K., 2017, ARA&A, 55, 389

Vogelsberger M., et al., 2014, Nature, 509, 177

Welsh L., Cooke R., Fumagalli M., Pettini M., 2022, ApJ, 929, 158

Welsh L., Cooke R., Fumagalli M., Pettini M., 2023, MNRAS, 525, 527

Werk J. K., et al., 2019, ApJ, 887, 89

Wolfe A. M., Gawiser E., Prochaska J. X., 2005, ARA&A, 43, 861

Yates R. M., Péroux C., Nelson D., 2021, MNRAS, 508, 3535

This paper has been typeset from a TEX/LATEX file prepared by the author.